\documentclass[journal]{IEEEtran}

\usepackage{amsmath,amssymb,amsfonts}
\usepackage{algorithmic}
\usepackage{textcomp}
\usepackage{times}
\usepackage{svg}
\usepackage{epsfig}
\usepackage{graphicx}
\usepackage{caption}
\usepackage{multirow}
\usepackage{blindtext}
\usepackage{booktabs}
\usepackage{cite}
\usepackage{algorithm, algorithmic}
\usepackage{soul}

\DeclareMathOperator*{\argmin}{arg\,min}

\usepackage[pagebackref=true,breaklinks=true,letterpaper=true,colorlinks,bookmarks=false]{hyperref}

\begin{document}
\title{GLEAM: Greedy Learning for Large-Scale Accelerated MRI Reconstruction}
\author{Batu Ozturkler, Arda Sahiner, Tolga Ergen, Arjun D Desai, Christopher M Sandino, Shreyas Vasanawala, John M Pauly \IEEEmembership{Senior Member, IEEE}, Morteza Mardani, Mert Pilanci \IEEEmembership{Member, IEEE}
\thanks{This work was supported by NIH Grants R01 EB009690, R01 EB026136 and NSF Grant DGE1656518.}
\thanks{Batu Ozturkler, Arda Sahiner, Tolga Ergen, Arjun D Desai, Christopher M Sandino, John M Pauly and Mert Pilanci are with the Electrical Engineering Department of Stanford University, 94305, USA. (ozt@stanford.edu, sahiner@stanford.edu, ergen@stanford.edu, arjundd@stanford.edu, sandino@stanford.edu, pauly@stanford.edu, pilanci@stanford.edu)}
\thanks{Morteza Mardani is with NVIDIA Inc, and the Electrical Engineering Department of Stanford University, 94305, USA. (morteza@stanford.edu)}
\thanks{Shreyas Vasanawala is with the Radiology Department of Stanford University, 94305, USA. (vasanawala@stanford.edu)}
}

\maketitle

\begin{abstract}
 Unrolled neural networks have recently achieved state-of-the-art accelerated MRI reconstruction. These networks unroll iterative optimization algorithms by alternating between physics-based consistency and neural-network based regularization. However, they require several iterations of a large neural network to handle high-dimensional imaging tasks such as 3D MRI. This limits traditional training algorithms based on backpropagation due to prohibitively large memory and compute requirements for calculating gradients and storing intermediate activations. To address this challenge, we propose Greedy LEarning for Accelerated MRI (GLEAM) reconstruction, an efficient training strategy for high-dimensional imaging settings. GLEAM splits the end-to-end network into decoupled network modules. Each module is optimized in a greedy manner with decoupled gradient updates, reducing the memory footprint during training. We show that the decoupled gradient updates can be performed in parallel on multiple graphical processing units (GPUs) to further reduce training time. We present  experiments with 2D and 3D datasets including multi-coil knee, brain, and dynamic cardiac cine MRI. We  observe that: $i)$ GLEAM generalizes as well as state-of-the-art memory-efficient baselines such as gradient checkpointing and invertible networks with the same memory footprint, but with 1.3x faster training; $ii)$ for the same memory footprint, GLEAM yields 1.1dB PSNR gain in 2D and 1.8 dB in 3D over end-to-end baselines.
 
\end{abstract}


\section{Introduction}
\label{sec:introduction}
\IEEEPARstart{M}{agnetic} resonance imaging (MRI) is a widely used medical imaging modality that provides high resolution information of the soft tissue anatomy, but is limited by long scan times.  Accelerated MRI reconstruction requires recovering high quality images from undersampled measurements. 
Deep learning reconstruction methods have recently gained popularity over parallel imaging and compressed sensing (CS) techniques \cite{sense,Lustig_Donoho_Pauly_2007} for accelerated MRI. A typical approach is model-based deep unrolling, where an iterative optimization algorithm is unrolled for a fixed number of iterations. The iterative algorithm alternates between enforcing data-consistency with the measurement model and a convolutional neural network (CNN) prior \cite{sandino2020compressed, Aggarwal_Mani_Jacob_2019, hammernik2018learning, mardani2018deep}. Unrolled networks are trained end-to-end in a supervised fashion with backpropagation (BP) where the parameters of the network are jointly optimized. In end-to-end training of unrolled networks with BP, the memory requirement increases linearly with the measurement count and the number of layers in the neural network. High-dimensional large-scale imaging applications such as Dynamic Contrast Enhanced (DCE) imaging, or blood flow imaging (4D-flow) have exceptionally large memory requirements, resulting in lower number of unrolled iterations which limits expressivity of unrolled networks trained with BP \cite{neuralprox,schlemper2017deep, nc-pdnet, CineSandino, KellmanMEL}. 

Therefore, reducing the memory footprint of training could enable the use of such networks in high-dimensional large-scale imaging applications where the memory requirements of a single measurement layer is large. In high-dimensional imaging where measurement layers have large memory requirements, a common method to reduce memory is to reduce the problem size. Some examples include coil compression, cropping to reduce spatial resolution, or reducing number of temporal frames in dynamic imaging \cite{CineSandino,schlemper2017deep}. In practice, in deep-learning based MRI reconstruction, it is beneficial for the model to have access to the full-scale imaging problem during training without the need to reduce problem size.

 \begin{figure*}[t!]
  \centering
  \begin{center}
      \includegraphics[width = 0.9\linewidth]{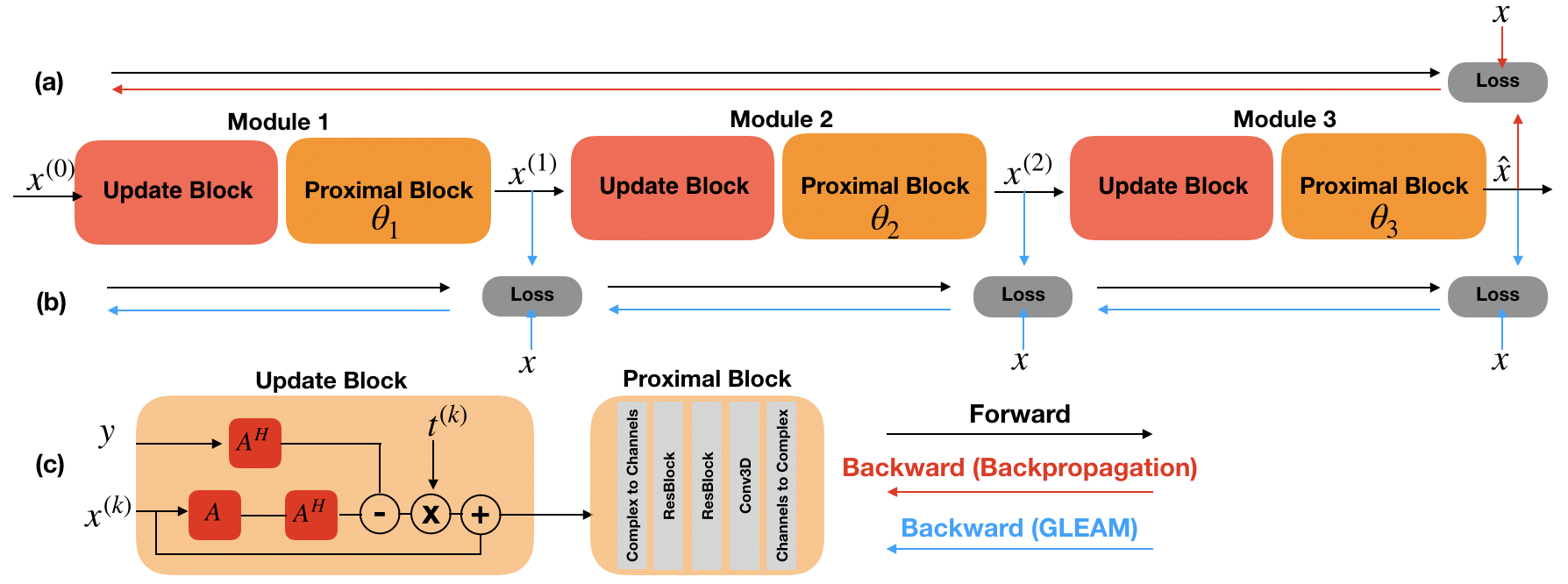}
  \end{center}
  \caption{\small \small Greedy learning for accelerated MRI (GLEAM) reconstruction. An example unrolled PGD network is shown with $N = M =3$ where $N$ is the number of unrolled iterations, $M$ is the number of modules. (a) End-to-end training with backpropagation. Forward pass is computed through all modules, and gradient updates are performed for all modules in the same backward pass. (b) GLEAM. At the end of each proximal block, loss is computed, and a local gradient update is performed on the current module. This procedure is repeated until the end of the network. (c) Update block consists of data-consistency with measurements, and the proximal block consists of residual blocks. 
  }
  \label{fig:gl-schematic}
\end{figure*}

 There exist several techniques to improve the memory footprint of unrolled networks. One such technique is gradient checkpointing \cite{checkpointing}, where a subset of activations are recomputed during backward pass, where the recomputation increases the overall computation time. Another line of work uses invertible networks to eliminate the need to store intermediate activations \cite{putzky2019invert, KellmanMEL, KeMEL}. However, the fixed-point algorithm used for inversion may be prone to numerical instabilities when a large number of layers are inverted \cite{KellmanMEL}.

In this work, we propose a greedy training objective for learning unrolled networks. We split the end-to-end network into decoupled network modules, and perform gradient updates on each module independently. Since each module has a significantly lower memory footprint, the memory requirement of the overall network decreases as well. Hence, our approach reduces memory when the number of measurements or number of layers is large. Furthermore, since our method obviates the need to calculate end-to-end gradients, the gradient updates of each module can be parallelized on multiple devices. Our contributions can be summarized as follows:
\begin{itemize}
    \item We propose GLEAM, a memory-efficient training strategy for unrolled MRI reconstruction via greedy learning.
    \item We empirically demonstrate that GLEAM generalizes as well as end-to-end learning and state-of-the-art memory-efficient baselines such as gradient checkpointing and invertible networks with the same memory footprint, but with $1.3\times$ faster training. 
    \item We also propose parallel GLEAM (P-GLEAM) which decouples and parallelizes gradient updates on multiple GPUs to further reduce training time by $1.6\times$.
    \item Given the same memory budget, our method yields 1.1dB PSNR gain in 2D, and 1.8dB in 3D over end-to-end baselines.
\end{itemize}

Our implementation is publicly available at \url{https://github.com/batuozt/GLEAM}. 
 
\section{Related Work}
\subsection{Memory-Efficient MRI Reconstruction}
Several works have aimed to tackle the memory bottleneck of training unrolled networks. These methods primarily differ by their memory-accuracy and memory-compute tradeoffs \cite{lowmemory}. A commonly used method with memory-compute tradeoff is gradient checkpointing (or activation checkpointing), where activations are stored for a subset of layers of the network instead of storing them for each layer as in BP \cite{checkpointing}. During backward pass, the activations that are omitted are recomputed. Thus, when activations are stored every $K$ layer for an $N$ layer network, the memory complexity reduces by a factor of $K$, where an additional forward pass needs to be performed for every omitted activation. Another line of work uses invertible networks where intermediate activations are calculated by reversing each layer, eliminating the need to store intermediate activations using BP \cite{putzky2019invert,KellmanMEL, KeMEL}. 

More recently, \cite{liu2021sgdnet} applied stochastic approximations to the data-consistency layers of unrolled networks to reduce memory complexity with increasing number of measurements, where the memory-accuracy tradeoff is favorable when approximate data consistency layers are sufficiently accurate. It must be noted that our method proposes a greedy optimization method that is compatible with any unrolled network. Thus, our method can be synergistically combined with any of these approaches to further reduce memory.

\subsection{Alternatives to end-to-end learning}
Recent works in computer vision have proposed alternatives to backpropagation (BP) for CNNs in the context of image classification. One approach is layer-wise training, where single hidden-layer subnetworks are trained sequentially until convergence while the previous part of the network remains frozen \cite{belilovsky2019greedy}. Another alternative is to use a decoupled learning approach, where a network is split into several smaller network modules, and gradient updates for each module are decoupled \cite{jaderberg2017decoupled,huo2018decoupled}, which was shown to achieve a similar generalization performance as BP \cite{belilovsky2020decoupled}. A crucial benefit of decoupled learning is achieving backward unlocking, which refers to updating network modules in parallel, without having to wait for other modules to provide gradients as in BP. 

Although memory efficiency is not the primary focus of those works, decoupled learning can improve the memory footprint of training by performing local gradient updates as opposed to end-to-end gradient updates in BP. 

\begin{figure*}[t]
  \centering
  \begin{center}
      \includegraphics[width = 0.9\linewidth]{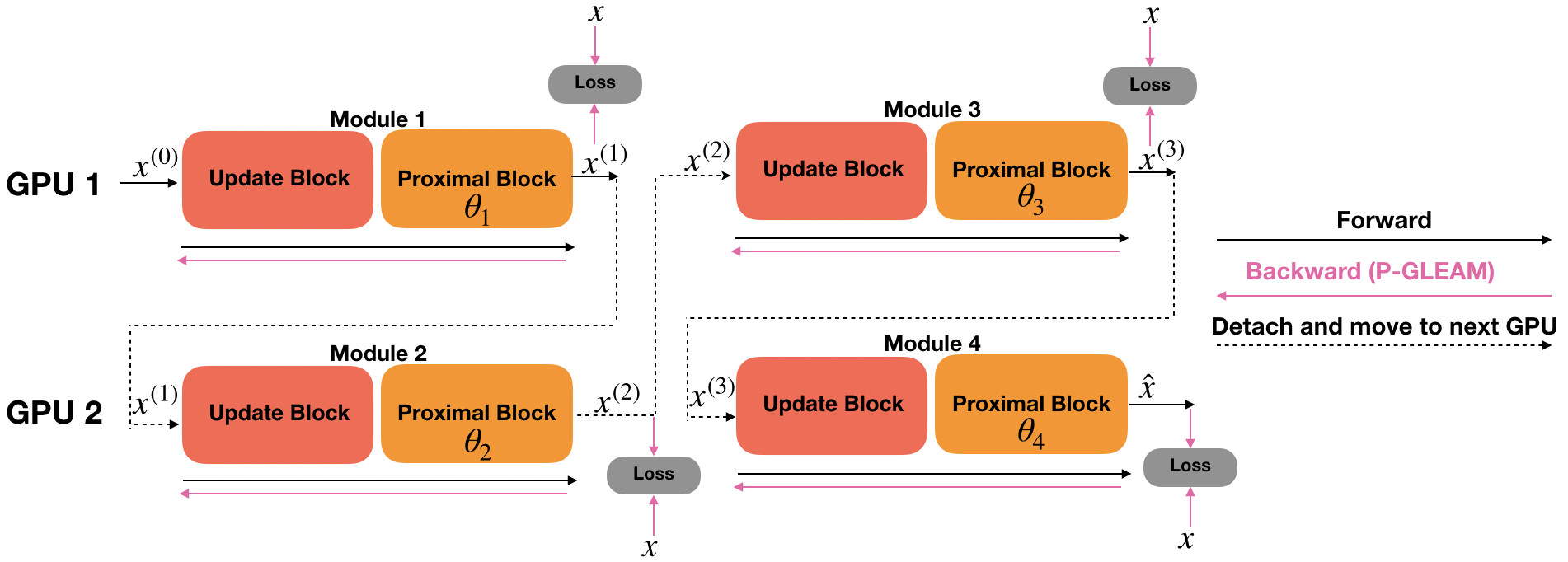}
  \end{center}
  \caption{\small \small Parallel GLEAM (P-GLEAM) for accelerated MRI reconstruction. An example unrolled PGD network with $B = 2$, $D = 2$, $M = 4$ is shown. The network weights are split across $D$ GPUs. When a mini-batch is sampled, it is passed to the network module on the first GPU. The forward pass is computed and the loss is calculated. Then, the result of the forward pass from the first module is detached, and sent to the network module on the second GPU, and a forward pass is computed on the second GPU. After forward pass is computed, backward pass is performed in parallel for both modules which is denoted by the pink arrows. This procedure is repeated until the end of the network. 
}
  \label{fig:adgl-schematic}
\end{figure*}

\section{Background}
\subsection{Accelerated MRI Reconstruction}
 The sensing model for accelerated MRI with parallel imaging and CS \cite{sense,Lustig_Donoho_Pauly_2007} can be expressed as
 \begin{equation}
     y =  \Omega FSx + \epsilon
 \end{equation}
 where $y$ denotes measurements in the Fourier domain (k-space), $x$ is the real image to reconstruct, $S$ are coil sensitivity maps associated with the receiver coils used in parallel imaging, $F$ is the Fourier transform matrix, $\Omega$ is the undersampling mask, and $\epsilon$ is additive noise. Then, the accelerated MRI reconstruction problem can be formulated as
 \begin{equation}
    \label{eq:forward}
     \hat{x} = \argmin_{x} \|Ax-y\|_{2}^{2}+\mu R(x)
 \end{equation}
 where $A =\Omega FS$ denotes the forward model, $R$ is a regularization function and $\mu$ is the regularization strength.
 
 \vspace{-0.35cm}
 
 \subsection{Proximal Gradient Descent (PGD)}
  A common choice to solve the optimization problem in Eq. \ref{eq:forward} is the proximal gradient descent (PGD) algorithm which alternates between two steps \cite{sandino2020compressed}
 \begin{equation}
     x^{(i+)} = \mathbf{DC}(x^{(i)}) := x^{(i-1)} - 2tA^{H}(A x^{(i)}-y)
 \end{equation}
 \begin{equation}
 \label{eq:step-2}
     x^{(i+1)} =  \mathbf{Prox}_R(x^{(i+)}) 
 \end{equation}
 where $i$ is the iteration number, $t$ is the step size, $A^{H}$ is the Hermitian transpose of the forward model operator A. $\mathbf{DC}$ denotes the  data-consistency operator, and $\mathbf{Prox}_R$ is the proximal operator. In unrolled PGD \cite{neuralprox,sandino2020compressed}, the proximal operator is learned with a CNN with trainable parameters. Then, \eqref{eq:step-2} can be rewritten as 
  \begin{equation}
 \label{eq:step-3}
     x^{(i+1)} =  f_{\theta_i}(x^{(i+)}) 
 \end{equation}
 where $f_{\theta_i}$ is the network, and $\theta_i$ its parameters at $i^{th}$ iteration.

\subsection{Model-Based Deep Learning (MoDL)}
Model-based deep learning (MoDL) is another alternative for accelerated MRI reconstruction which proposes to solve the following optimization problem:
\begin{equation}
\label{eq:modl}
     \hat{x} = \argmin_{x}\|Ax-y\|_{2}^{2}+\mu\|x-R(x)\|^{2}
 \end{equation}
 In MoDL, the problem in Eq. \ref{eq:modl} is solved using the half quadratic splitting (HQS) algorithm in alternating manner \cite{Aggarwal_Mani_Jacob_2019}. Instead of performing a single gradient update as in PGD, HQS performs a full model inversion. The update rule for MoDL is as follows
\begin{equation}
\label{eq:modl-step1}
\begin{split}
     x^{(i+1)} &= \argmin_{x}\|Ax-y\|_{2}^{2}+\mu\|x-z^{(i)}\|^{2} \\
     &= (A^HA+\mu I)^{-1}(A^Hy + \mu z^{(i)}) \\
\end{split}
\end{equation}
\begin{equation}
\label{eq:modl-step2}
z^{(i)} = R(x^{(i)}) =  f_{\theta_i}(x^{(i)}) 
\end{equation}
where Eq. \ref{eq:modl-step1} can be solved efficiently using the conjugate gradient (CG) method which performs the full model inversion, and the regularization function $R$ is learned with a CNN denoted as $f_{\theta_i}$ at the $i^{th}$ iteration.

\vspace{-0.2cm}

\subsection{End-to-End Learning}
In unrolled networks, end-to-end learning is employed by unrolling either of the iterative optimization algorithms described above for a fixed number of iterations $N$. Such a models parameters can be jointly optimized using BP \begin{equation}
 \label{eq:e2e-loss}
     \min_{\theta_{1},..,\theta_{N}}\sum_{k}\mathcal{L}(f_{\theta_{1},..,\theta_{N}}(y_{k},A_{k}),x_{k})
 \end{equation}
 where $f_{\theta_{1},..,\theta_{N}}$ is the end-to-end network, $y_k$ is the $k^{th}$ undersampled image, $x_{k}$ is the $k^{th}$ fully-sampled reference, $\mathcal{L}$ is the loss function.
\section{Methods}
\subsection{Greedy LEarning for Accelerated MRI Reconstruction (GLEAM)}
In this section, we describe our method for optimizing unrolled networks. Here, we describe GLEAM for unrolled PGD, although an extension to MoDL is straightforward, where the only difference is replacing the gradient-based data-consistency update with the conjugate gradient (CG) algorithm. In GLEAM, we extend the decoupled greedy learning approach in \cite{belilovsky2020decoupled} for training CNNs to the setting of unrolled networks. We consider an unrolled network with $N$ number of unrolled iterations, where the network is split into $M$ modules, such that $\frac{N}{M}$ is the number of unrolled iterations in each module. For simplicity, we outline the case where $N = M$. In particular, let $x^{(0)} = y$ be the input, $x$ the fully sampled reference, $x^{(m)} = f_{\theta_{m}}(\mathbf{DC}(x^{(m-1)}))$ the output of the $m^{th}$ network module, and $S$ the number of mini-batches. For the $m^{th}$ network module, let the cost function be $\mathcal{L}(f_{\theta_{m}}(\mathbf{DC}(x^{(m-1)})),x)$. Then, the parameters of module $m$ can be optimized using the greedy training objective:
  \begin{equation}
 \label{eq:greedy-loss}
     \min_{\theta_{m}}\mathcal{L}(f_{\theta_{m}}(\mathbf{DC}(x^{(m-1)})),x).
 \end{equation}
 Our method to optimize the greedy training objective in Eq. \ref{eq:greedy-loss} for $M$ modules is outlined in Alg. \ref{alg:gr-alg}. When a mini-batch is sampled during training, the first module computes the forward pass, then performs a gradient update based on its own local loss. After the gradient update, the result of the forward pass is sent to the next module, and the activations for the current module are discarded from GPU memory. This procedure is repeated until the final module, which produces the final prediction of the network, after which a new mini-batch is sampled. Individual updates of each set of parameters are performed independently across different modules, and computations are only performed for one module at every instant. Thus, the memory requirement scales with the size of each module, rather than the size of the end-to-end network, which drastically reduces the memory requirement. In our approach, we update each module using BP. The unrolled network trained with GLEAM is illustrated in Fig. \ref{fig:gl-schematic}.

\begin{algorithm}[t!]
 \caption{Proximal Gradient Descent with GLEAM}
 \label{alg:gr-alg}
 \begin{algorithmic}[1]
 \renewcommand{\algorithmicrequire}{\textbf{Input:}}
 \REQUIRE Mini-batches ${\{(x_k^{(0)},x_k)\}_{k \leq S}}$
 \\ \textit{Initialize} : Parameters $\{\theta_m\}_{m \leq M}$
  \FOR {$k \in 1,...,S$}
  \FOR {$m \in 1,...,M$}
  \STATE $x_k^{(m)} \gets f_{\theta_{m}}(\mathbf{DC}(x_k^{(m-1)}))$
  \STATE Compute $\nabla_{\theta_m}\mathcal{L}(x_k^{(m)},x_k)$
   \STATE $\theta_m \gets \textnormal{Update Parameters}(\theta_m)$.
  \ENDFOR
  \ENDFOR
 \end{algorithmic} 
 \end{algorithm}

\begin{algorithm}[t!]
 \caption{Proximal Gradient Descent with P-GLEAM}
 \label{alg:async-gr-alg}
 \begin{algorithmic}[1]
 \renewcommand{\algorithmicrequire}{\textbf{Input:}}
 \REQUIRE Mini-batches ${\{(x_k^{(0)},x_k)\}_{k \leq S}}$ on GPU $d=1$  
 \\ \textit{Initialize} : Parameters $\{\theta_m\}_{m \leq M}$
  \FOR {$k \in 1,...,S$}
  \FOR {$b \in 1,...,B$}
  \FOR {$d \in 1,...,D$}
  \STATE $m = (b-1)*D+d$
  \STATE $x_k^{(m)} \gets f_{\theta_{m}}(\mathbf{DC}(x_k^{(m-1)}))$
  \STATE Compute $\mathcal{L}(x_k^{(m)},x_k)$  \label{eq:compute_loss}
  \STATE Detach $x_k^{(m)}$ and move to $(d+1)^{th}$ GPU
  \ENDFOR
  \FOR {$d \in 1,...,D$}
  \STATE $m = (b-1)*D+d$
  \STATE Compute $\nabla_{\theta_m}\mathcal{L}(x_k^{(m)},x_k)$  \label{eq:compute_grad}
  \STATE $\theta_m \gets \textnormal{Update Parameters}(\theta_m)$. \label{eq:update_parameters}
  \ENDFOR
  \ENDFOR
  \ENDFOR
 \end{algorithmic} 
 \end{algorithm}

\vspace{-0.2cm}

\subsection{Parallel GLEAM (P-GLEAM)}
We are now ready to extend GLEAM to a distributed setting with $D$ GPUs. Since gradient updates for each module are independent, we can distribute modules across GPUs to perform parallel training. We consider the case where the end-to-end network is split to $M$ modules to perform Parallel GLEAM (P-GLEAM) with $N = M$. We seek to parallelize the computations across $D$ GPUs, and accordingly separate our $M$ modules into $B := \frac{M}{D}$ blocks, such that each module in one block has its computation performed on a unique GPU. 

P-GLEAM is outlined in Alg. \ref{alg:async-gr-alg}. When a mini batch is sampled, it is passed to the network module on the first GPU. The forward pass is computed and the loss is calculated. Then, the result of the forward pass from the first module is detached, and sent to the second module on the second GPU, and a forward pass is computed on this GPU. This procedure is repeated until the last GPU on the first block. After all of the forward passes for the first block are computed, gradient updates for all modules on the first block can be performed in parallel on each individual GPU. This procedure is repeated on all $B$ blocks until the end of the network. The unrolled network trained with P-GLEAM is illustrated in Fig. \ref{fig:adgl-schematic}.

\begin{figure*}[t!]
  \centering
  \begin{center}
      \includegraphics[width = 0.9\linewidth]{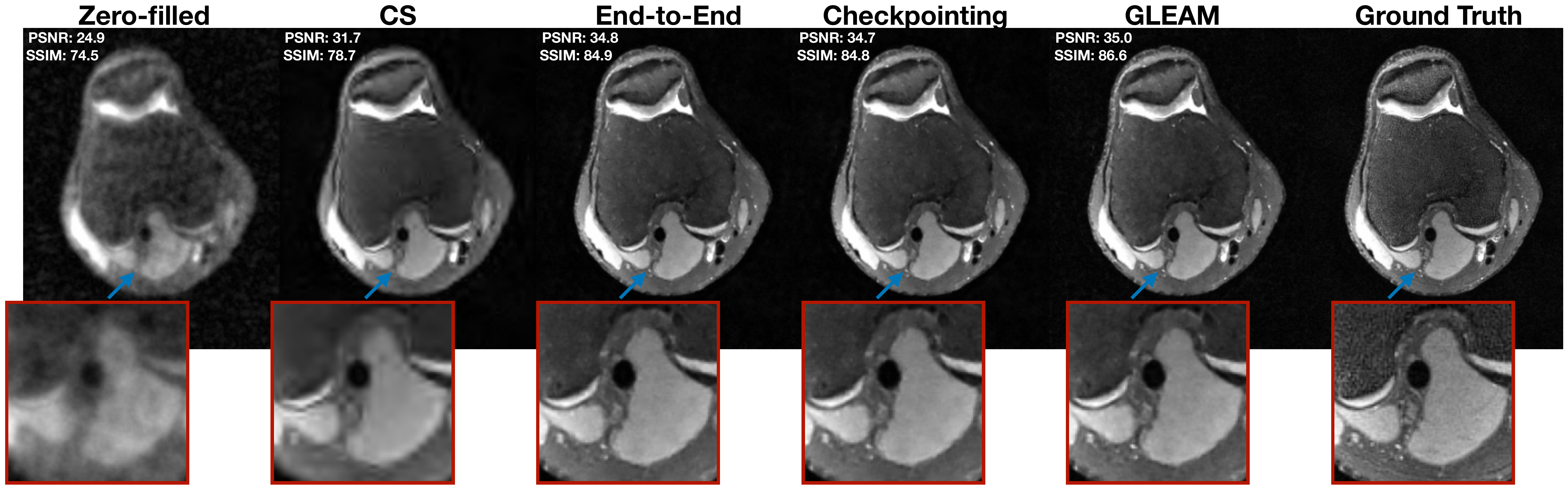}
  \end{center}
  \caption{\small{Reconstruction of an example knee slice from the Mridata test set with different reconstruction methods at $R = 16$ using 2D MoDL. GLEAM produces similar visual quality with end-to-end learning and checkpointing while achieving low memory and fast training time. Reconstructed images were normalized by the $99^{th}$ percentile of the image magnitude for better dynamic range.}}
  \label{fig:knee-mr}
\end{figure*}

In the P-GLEAM setup in Alg. \ref{alg:async-gr-alg}, for each GPU, once the first forward pass is performed in \eqref{eq:compute_loss}, the backward pass computations in \eqref{eq:compute_grad} and \eqref{eq:update_parameters} can begin. After all forward passes have been computed, the backward passes that are now distributed on $D$ GPUs can be computed in parallel. In practice, this can be implemented by utilizing the asynchronous nature of GPU operations in CUDA, where BP for modules on different devices is performed asynchronously \cite{CUDA,paszke2019pytorch}.

Both GLEAM and P-GLEAM compute the same local loss, where the computations that are performed on one device in GLEAM are distributed on multiple devices for P-GLEAM. As a result, the gradient updates for GLEAM and P-GLEAM are numerically equivalent.

Among methods to parallelize deep networks, two popular choices are data parallelism \cite{goyal2017accurate} and model parallelism \cite{modelparallel}. In data parallelism different input mini-batches are split across multiple GPUs, whereas in model parallelism the features of convolutional layers are split across multiple GPUs. In contrast, the parallelism in P-GLEAM is similar to layer-wise or module-wise parallelism \cite{huo2018decoupled,belilovsky2020decoupled,jaderberg2017decoupled} where gradient calculations for different network modules are parallelized, achieving backward unlocking. Thus, P-GLEAM can be combined with data parallelism and model parallelism for distributed training.

\vspace{-0.4cm}

\section{Experiments}
In these experiments, we seek to answer two questions: 
\begin{itemize}
    \item Given the same network architecture, can GLEAM achieve a lower memory footprint and better training time compared to baselines?
    \item Given the same hardware with a fixed memory constraint, can GLEAM allow for more learnable parameters and improve reconstruction quality?
\end{itemize}
We show that GLEAM $i)$ generalizes as well as state-of-the-art memory-efficient baselines with the same memory footprint and faster training, and $ii)$ for the same memory footprint, improves reconstruction over end-to-end baselines. To illustrate the applicability of our method to various architectures, we demonstrate it on unrolled PGD \cite{sandino2020compressed} and MoDL \cite{Aggarwal_Mani_Jacob_2019}.

\subsection{Datasets}
\label{sec:dataset}
\subsubsection{Mridata Multi-Coil Knee Dataset}
We used the fully-sampled 3D fast-spin echo (FSE) multi-coil knee MRI dataset publicly available on mridata.org \cite{ong2018Mridata}. Each 3D volume had a matrix size of $320 \times 320 \times 256$ with $8$ coils, where training was performed 2D by treating each axial slice with size $320 \times 256$ as separate examples. The dataset consisted of unique scans from $19$ subjects, where $14$ subjects ($4480$ slices) were used for training, $2$ subjects ($640$ slices) were used for validation, and $3$ subjects ($960$ slices) were used for testing. Sensitivity map estimation was performed in SigPy \cite{sigpy} using JSENSE \cite{jsense} with kernel width of 8 for each volume. Fully-sampled references were retrospectively undersampled using a 2D Poisson Disc undersampling mask at $R = \{4,12,16\}$, where $R$ denotes acceleration factor. 
\subsubsection{FastMRI Brain Multi-Coil Dataset}
We used the multi-coil brain MRI dataset from fastMRI \cite{zbontar2018fastmri}. All data splits were filtered to include only T2-weighted scans acquired at a 3T field strength to control for confounding variables. From this data split, 54 scans ($858$ slices) were used for training, and $50$ scans were used for validation ($790$ slices) and $50$ scans ($800$ slices) were used for testing where the matrix size for each scan was $384 \times 384$ with $15$ coils. Sensitivity maps were estimated using ESPIRiT with a kernel width of $8$ and a calibration region of $12 \times 12$ \cite{uecker2014espirit}. At acceleration factor $R = 8$, this calibration region is $4\%$ of the autocalibration region.
Fully-sampled references were retrospectively undersampled using 1D random undersampling, a pattern widely used in 2D Cartesian imaging \cite{Lustig_Donoho_Pauly_2007}.

\begin{figure*}[t!]
  \centering
  \begin{center}
      \includegraphics[width = 0.9\linewidth]{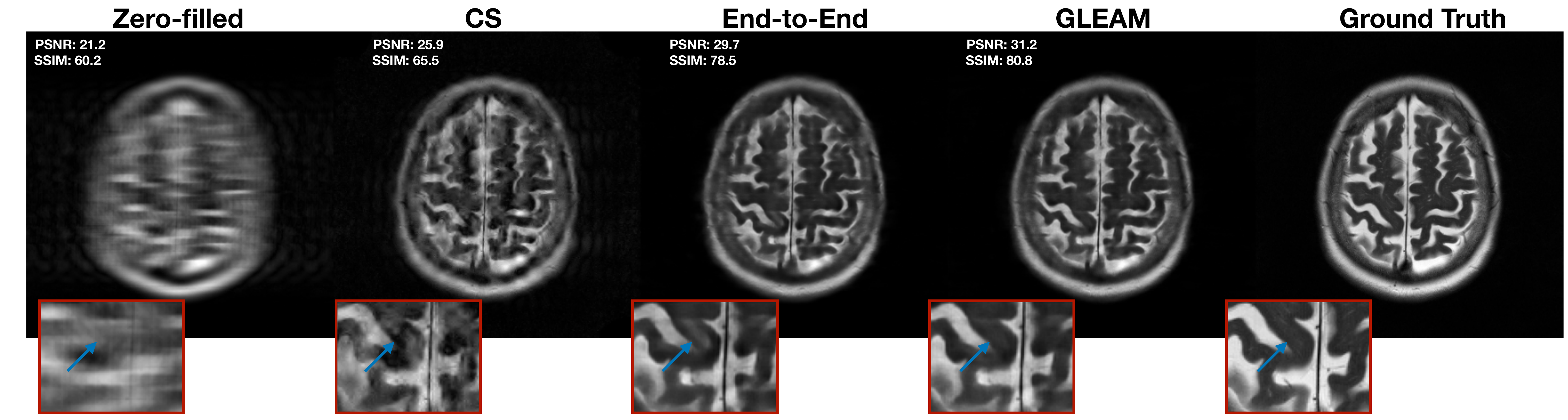}
  \end{center}
  \caption{\small{Reconstruction of an example brain slice from the fastMRI test set with different reconstruction methods at $R = 8$ using 2D unrolled PGD. Under the same memory budget, GLEAM improves visual quality and better mitigates aliasing artifacts. Reconstructed images were normalized by the $99^{th}$ percentile of the image magnitude for better dynamic range.
  }}
  \label{fig:fastmri-pgd}
\end{figure*}

\subsubsection{Dynamic Cardiac Cine MRI}
We adopted the fully-sampled balanced SSFP 2D+time cardiac cine dataset used in \cite{sandino2021deep} for evaluation, which is a larger version of the dataset in \cite{CineSandino}. Training scans from $17$ subjects were split slice-by-slice to create $261$ unique cine slices. For validation, $2$ subjects ($33$ slices) were used, and for testing, $5$ subjects ($76$ slices) were used where each cine slice had $20$ temporal frames. The matrix size for each scan varied within the range from $144-216 \times 180-242$ with $8$ coils. Scans were retrospectively undersampled using a k-t variable-density mask at $R = 10$.

\vspace{-0.3cm}

\subsection{Baseline Methods}
\subsubsection{End-to-End Learning}
We employed 2D \cite{sandino2020compressed} and 3D \cite{CineSandino} unrolled PGD, and 2D and 3D MoDL \cite{Aggarwal_Mani_Jacob_2019} as end-to-end learning baselines.
\subsubsection{Compressed Sensing (CS)}
We implemented CS with $l_1$-wavelet regularization \cite{Lustig_Donoho_Pauly_2007}. Reconstruction was performed slice-by-slice using SigPy where the proximal gradient method was run for 100 iterations \cite{sigpy}. The regularization strength was visually tuned for different datasets and acceleration factors. The regularization was determined to be $\{0.12,0.07,0.06\}$ for Mridata at $R = \{4,12,16\}$ and $0.001$ for fastMRI at $R = 8$.

For CS dynamic cardiac cine MRI, we used a $l_1$-wavelet regularized reconstruction method with
spatial and temporal total variation (TV) regularization \cite{uecker2014espirit}. Reconstruction was performed for each cine slice using SigPy where the proximal gradient method was run for 200 iterations \cite{sigpy}. Regularization strengths for
spatial and temporal TV priors were $0.05$ and $0.01$ respectively for the cardiac cine dataset at $R = 10$, which were determined with visual tuning.

\subsubsection{Memory-Efficient Learning}
\begin{itemize}
    \item Gradient Checkpointing: We implemented gradient checkpointing \cite{checkpointing} for MoDL. A checkpoint was placed at the end of each unrolled iteration, where number of checkpoints $N_{cp}$ was equal to the number of iterations $N$. 
    \item Invertible Learning: We implemented invertible learning applied to MoDL from \cite{KellmanMEL} and \cite{KeMEL}, termed memory-efficient learning (MEL). In MEL, an invertible CNN is incorporated to the regularizer in Eq. \ref{eq:modl-step2}, and it is inverted using a fixed-point algorithm \cite{fixedpoint}. On the other hand, the CG step in Eq. \ref{eq:modl-step2} is inverted by
    \begin{equation}
    \label{eq:modl-inverse}
     z^{(i)} = \frac{1}{\mu}((A^HA+\mu I)x^{(i+1)}-A^Hy) \\
    \end{equation}
    As discussed in \cite{KellmanMEL}, the accuracy of the inversion in Eq. \ref{eq:modl-inverse} is ensured only if CG can perform the model inversion in Eq. \ref{eq:modl-step2} accurately. To mitigate this source of numerical error, a hybrid reverse recalculation and checkpointing method is used in \cite{KellmanMEL}. In our implementation of MEL, we carefully tuned the number of checkpoints $N_{cp}$ to ensure numerical stability. Thus, $N_{cp} = 7$ was used in all MEL experiments. As a general rule, $N_{cp}$ should be increased until numerical stability is guaranteed.
\end{itemize}

\vspace{-0.4cm}

\subsection{Implementation Details}

Architecture details for experiments with each unrolled algorithm and dataset are described below. In all experiments, the end-to-end architecture with most number of unrolled iterations, and number of features that could fit on a 12 GB GPU were selected for end-to-end baselines. 

\begin{figure*}[t!]
  \centering
  \begin{center}
      \includegraphics[width = 0.9\linewidth]{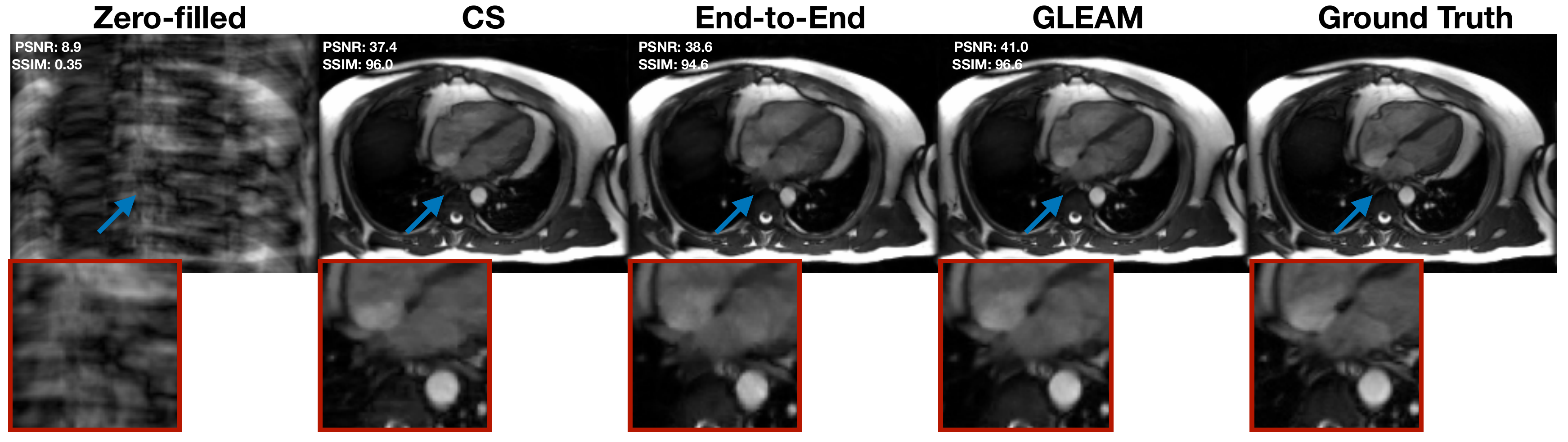}
  \end{center}
  \caption{\small{Reconstruction of an example slice from the test set of the cardiac cine dataset with different reconstruction methods at $R = 10$ using 3D unrolled PGD. Under the same memory budget, GLEAM improves visual quality and recovers fine structures. Reconstructed images were normalized by the $95^{th}$ percentile of the image magnitude for better dynamic range.
  }}
  \label{fig:cine-pgd}
\end{figure*}

\subsubsection{PGD}

\begin{itemize}
    \item 2D: The unrolled PGD network from \cite{sandino2020compressed} was adopted where each proximal block consisted of 2 residual blocks. Each residual block consisted of 2 ReLUs followed by convolutional layers with no normalization. For Mridata and fastMRI, a batch size of $4$ and $2$ were used and networks were trained for 50000 and 80000 iterations, respectively. Both networks had $N_f = 100$ where $N_f$ denotes number of features. The end-to-end network had $N = 8$ unrolled iterations.
    \item 3D: The unrolled PGD network proposed by \cite{CineSandino} with 3D convolutions was used. The network had $N_f = 60$, $N = 4$, a batch size of 1, and was trained for 25000 iterations.
\end{itemize}

\subsubsection{MoDL}

\begin{itemize}
    \item 2D: The network from \cite{Aggarwal_Mani_Jacob_2019} was adopted. For fair performance comparison with MEL, all reconstruction methods used invertible residual CNNs \cite{gomez2017reversible}. The residual CNN included 5 convolutional layers followed by ReLU activations with one skip connection and no normalization as in \cite{KeMEL}. As the numerical stability of MEL was found to be sensitive to the initial value of the regularization parameter, we searched for values of $\mu$ that make MEL converge. As a result, $\mu = \{4.0, 0.05\}$ were used for $R = \{4, 16\}$ respectively. The network had 10 CG steps, $N_f = 70$, $N = 8$, a batch size of $4$, and was run for 50000 iterations. 
    \item 3D: The network from the 2D setup was used with 3D convolutions with a kernel size of $3 \times 3 \times 3$. The network had 10 CG steps with $\mu = 0.05$, $N_f = 50$, $N = 4$, a batch size of $1$, and was run for 25000 iterations. 
\end{itemize}

\subsubsection{GLEAM}
In the experiments where the architecture was fixed for all methods, GLEAM used $N = M$. On the other hand, in the experiments where memory was fixed for all methods, GLEAM used a $3 \times$ deeper network. Thus, for fastMRI, GLEAM used $N = 24$ with $M = 3$ where the end-to-end network used $N = 8$, and for cardiac cine PGD, GLEAM used $N = 12$ with $M = 3$ where the end-to-end network used $N = 4$.

\subsubsection{Training}
Networks were trained with zero-filled, SENSE-reconstructed complex images generated using the estimated sensitivity maps with a complex $l_1$ loss. Complex images were represented by their real and imaginary components with two-channels. All end-to-end networks, as well as each network module in GLEAM were trained with the Adam optimizer with $\beta_1 = 0.9$, $\beta_2 = 0.999$ \cite{kingma2014adam}. At inference time, the model checkpoint that achieved the lowest validation nRMSE was selected for each method. 

All experiments were conducted using Pytorch \cite{paszke2019pytorch} on a single Geforce GTX Titan Xp with 12 GB memory. For evaluation, normalized root-mean-square error (nRMSE), structural similarity (SSIM), and peak signal-to-noise ratio (PSNR, dB) were used. Metrics were computed for each test slice over magnitude images.

\begin{table}[t!]
\centering
 \resizebox{\columnwidth}{!}{
\begin{tabular}{llccccc}
\toprule

\textit{R} & Method &      SSIM &          nRMSE &       PSNR   &Memory & Timing\\ 
& & & & (dB) & (GB) & (iter/sec) \\
\midrule

\multirow{3}{*}{12x} 
    & End-to-End & 0.889(0.006) & 0.127(0.007) & 40.03(0.33)&11.1& 1.39\\
    & GLEAM & \textbf{0.913(0.002)} & 0.127(0.007) & 40.09(0.34) &\textbf{2.0}& 1.38\\
    & CS & 0.846(0.010) & 0.175(0.010)& 37.28(0.27)&N/A&N/A\\

\midrule

\multirow{3}{*}{16x} 
    & End-to-End & 0.892(0.006) & 0.133(0.008) & 39.66(0.33)&11.1& 1.39\\
    & GLEAM & \textbf{0.899(0.004)} & 0.135(0.008) & 39.55(0.33)&\textbf{2.0}& 1.38\\
    & CS &0.847(0.009)&0.178(0.011)&37.12(0.28)&N/A&N/A\\
    
\bottomrule
\end{tabular}
}
\caption{\small{Reconstruction methods applied to the 2D unrolled PGD architecture at $R = \{12,16\}$ on Mridata. Comparison of image metrics, maximum GPU memory during training and training time per iteration. Metrics are reported as \textit{mean (standard deviation)}.}} 
\label{tbl:2dpgd}
\end{table}

\vspace{-0.5cm}

\section{Results}
\label{sec:results}

\subsection{Mridata}
\subsubsection{2D PGD}
Table \ref{tbl:2dpgd} shows average performance of reconstruction methods, maximum GPU memory during training and training time per iteration at $R = \{12, 16\}$ on the Mridata test set. GLEAM achieved  on par performance with end-to-end learning, while achieving $5.6 \times$ less memory and similar training time with end-to-end learning. Both deep-learning reconstruction methods outperformed CS.

\subsubsection{2D MoDL}
Table \ref{tbl:2d-modl} shows average performance of reconstruction methods, maximum GPU memory during training and training time per iteration at $R = \{4, 16\}$ on the Mridata test set. GLEAM achieved on par performance with end-to-end learning, gradient checkpointing, and MEL across all performance metrics. Furthermore, it had $6.8 \times$ less memory than end-to-end learning with $1.4 \times$ faster per iteration training time compared to gradient checkpointing and MEL. In addition, GLEAM achieved slightly less memory compared to gradient checkpointing since no additional activations are stored for gradient calculation. MEL did not converge at $R = 16$. Example knee slices are illustrated in Fig. \ref{fig:knee-mr}.

\vspace{-0.2cm}

\subsection{FastMRI}
\subsubsection{2D PGD}
GLEAM allowed for training a network with $3 \times$ as many unrolled iterations as end-to-end learning. As a result, the network trained with GLEAM outperformed the network trained with end-to-end learning by $+1.1$ dB PSNR, $+0.016$ SSIM and $+0.020$ nRMSE (Table \ref{tbl:fastmri-prox2d}) with the same memory footprint. Example knee slices from the test set are shown in Fig. \ref{fig:fastmri-pgd}. GLEAM better mitigated aliasing artifacts compared to end-to-end learning, improving reconstruction quality. 

\vspace{-0.2cm}

\subsection{Dynamic Cardiac Cine MRI}
\subsubsection{3D PGD}
GLEAM allowed for training a network with $3 \times$ as many unrolled iterations as end-to-end learning. As a result, the network trained with GLEAM outperformed the network trained with end-to-end learning by $+1.8$ dB PSNR, $+0.003$ SSIM and $+0.013$ nRMSE (Table \ref{tbl:cine3D}) with the same memory footprint. Example cardiac cine slices from the test set are shown in Fig. \ref{fig:cine-pgd}. GLEAM improved reconstruction quality compared to end-to-end learning and recovered fine anatomical structures. 

\begin{figure*}[t!]
  \centering
  \begin{center}
      \includegraphics[width = 0.9\linewidth]{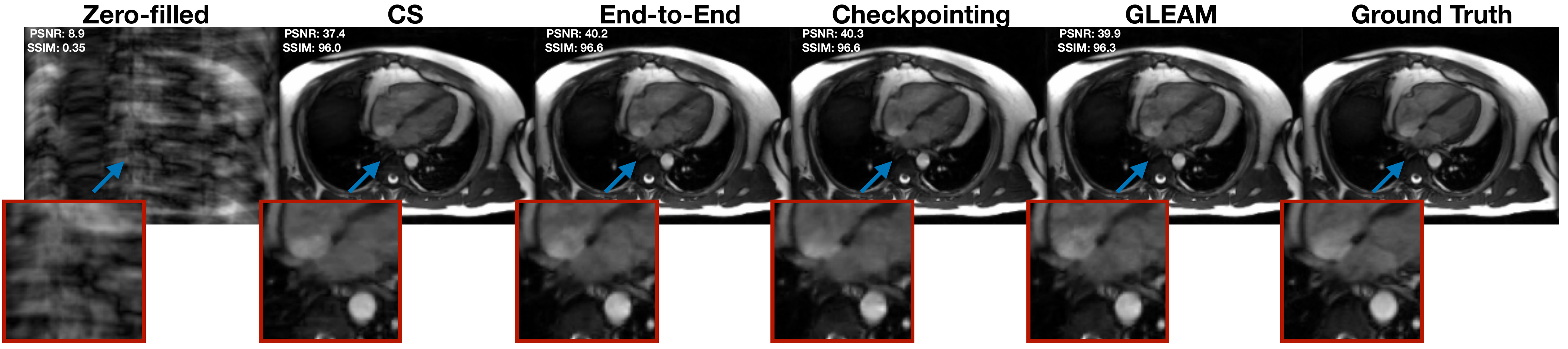}
  \end{center}
  \caption{\small{Reconstruction of an example slice from the test set of the cardiac cine dataset with different reconstruction methods at $R = 10$ using 3D MoDL. GLEAM produces similar visual quality with end-to-end learning and checkpointing while achieving low memory and fast training time. Reconstructed images were normalized by the $95^{th}$ percentile of the image magnitude for better dynamic range. 
  }}
  \label{fig:cine-modl}
\end{figure*}

\subsubsection{3D MoDL}
Table \ref{tbl:3d-modl} shows average performance of reconstruction methods, maximum GPU memory during training and training time per iteration at $R = 10$ on the test set of the cardiac cine dataset. GLEAM performed on par with end-to-end learning and gradient checkpointing. Furthermore, it had $3.6 \times$ less memory than end-to-end learning, and $1.3 \times$ faster per iteration training time compared to gradient checkpointing. MEL did not converge for this application. Example cardiac cine slices are illustrated in Fig. \ref{fig:cine-modl}.

\begin{table}[t!]
\centering
 \resizebox{\columnwidth}{!}{
\begin{tabular}{llccccc}
\toprule

\textit{R} & Method &      SSIM &          nRMSE &       PSNR   &Memory & Timing\\ 
& & & & (dB) & (GB) & (iter/sec) \\
\midrule

\multirow{5}{*}{4x} 
    & End-to-End &  0.934(0.003) & 0.098(0.006) & 42.30(0.49) & 11.4 & \textbf{2.09} \\
     & Checkpointing & 0.934(0.003) & 0.099(0.006) & 42.26(0.49) & 1.67 & 2.83 \\
     & MEL & 0.937(0.003) & 0.097(0.006) & 42.36(0.47) & 1.67 & 2.96 \\
     & GLEAM & \textbf{0.941(0.005)} & 0.098(0.007) & 42.30(0.50) & \textbf{1.57} & \textbf{2.08}\\
     & CS &0.841(0.011) & 0.167(0.008) & 37.69(0.19) & N/A & N/A\\

\midrule

\multirow{4}{*}{16x} 
     & End-to-End & 0.893(0.002) & 0.134(0.009) & 39.59(0.42) & 11.4 & \textbf{2.09} \\
     & Checkpointing & 0.893(0.002) & 0.134(0.009) & 39.58(0.42) & 1.67 & 2.83 \\
     & MEL & --- & --- & --- & 1.67 & 2.96 \\
     & GLEAM & \textbf{0.922(0.008)} & 0.133(0.009) & 39.65(0.45) & \textbf{1.57} & \textbf{2.08}\\
     & CS &0.847(0.009)&0.178(0.011)&37.12(0.28) & N/A & N/A\\
    
\bottomrule
\end{tabular}
}
\caption{\small \small {Reconstruction methods applied to the 2D MoDL architecture at $R = \{4,16\}$ on Mridata. Comparison of image metrics, maximum GPU memory during training and training time per iteration. MEL did not converge at $R = 16$.}} 
\label{tbl:2d-modl}
\end{table}

\begin{table}[t!]
\centering
 \resizebox{\columnwidth}{!}{
\begin{tabular}{llccccc}
\toprule

 Method &      SSIM &          nRMSE &       PSNR   &  Memory & Parameters &   \\
 &&&(dB)&(GB)&\\
\midrule

     End-to-End & 0.854(0.036) & 0.182(0.014) & 31.09(1.13)&9.9& 2.19M \\
     GLEAM & \textbf{0.870(0.033)} & \textbf{0.162(0.015)} & \textbf{32.14(1.22)}&10.2& \textbf{6.57M} \\
     CS & 0.716(0.055) & 0.299(0.065) & 26.93(1.90)&N/A&N/A \\
\bottomrule
\end{tabular}
}
\caption{\small \small{Reconstruction methods applied to the 2D unrolled PGD architecture at $R = 8$ on fastMRI. Comparison of image metrics, maximum GPU memory during training and number of learnable parameters for each method.}}
\label{tbl:fastmri-prox2d}
\end{table}

\vspace{-0.4cm}

\subsection{Timing}
\label{sec:timing}
\begin{figure}[t!]
  \centering
  \begin{center}
      \includegraphics[width = 0.75\linewidth]{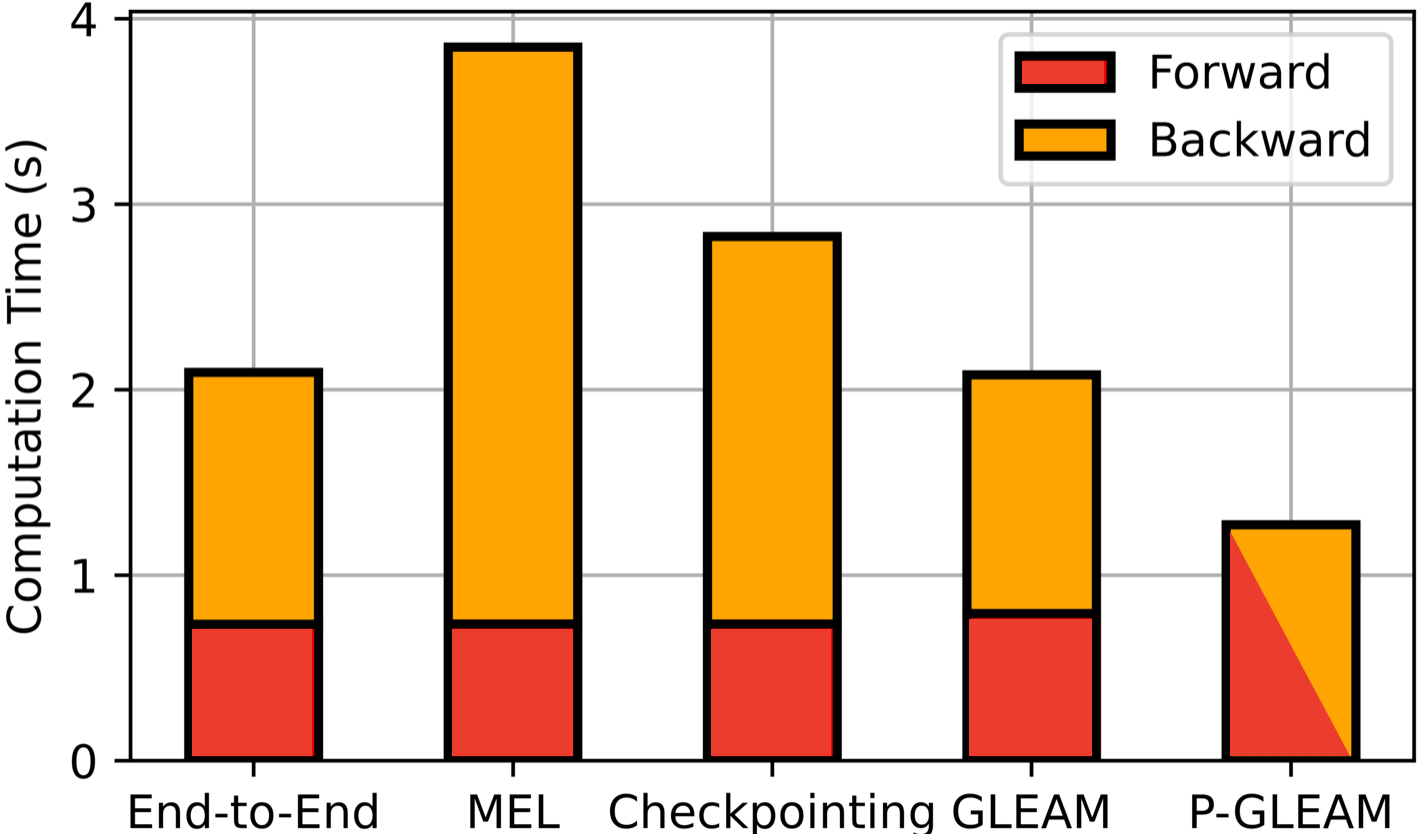}
  \end{center}
  \caption{\small{Forward and backward time per iteration during training. Methods were benchmarked on 2D MoDL on the same hardware. MEL and checkpointing are slower than end-to-end learning due to additional computations during backward pass. On a single GPU, BP and GLEAM have similar computation time. When modules are split on 2 GPUs in P-GLEAM, backward time is reduced to speed up training. In P-GLEAM, forward and backward time cannot be computed in isolation since forward and backward passes are computed in parallel.
  }}
  \label{fig:timing}
\end{figure}

Let the time for forward pass through one unrolled layer be $T_f$, the time for backward pass for N unrolled layers be $T^N_b$, and the time for performing the backward pass by inverting an unrolled layer for MEL be $T_{inv}$. Fig. \ref{fig:timing} shows forward and backward time comparison for reconstruction methods per iteration during training for the 2D MoDL network. In this comparison, MEL does not use any checkpoints and uses inversion to calculate gradients for each layer to isolate the impact of $T_{inv}$ and $T_f$ on training time. In our application, the time to invert layers using the fixed-point algorithm with MEL ($T_{inv}$) takes longer than recomputing the forward pass in gradient checkpointing ($T_f$), as a result MEL has a longer backward time where $T_{inv} > T_{f}$. In general, the comparison between $T_f$ and $T_{inv}$ varies by application \cite{KellmanMEL}. 

The forward time for GLEAM is slightly larger due to the time required to detach the activations from one module and pass it to the next module. Conversely, the backward time of GLEAM is slightly faster, since backward time in end-to-end learning ($T^N_b$) takes longer than that of GLEAM ($N*(T^1_b)$) in our application ($T^N_b>N*(T^1_b)$). Thus, end-to-end learning and GLEAM have similar overall computation time.

\begin{table}[t!]
\centering
 \resizebox{\columnwidth}{!}{
\begin{tabular}{llccccc}
\toprule

 Method & SSIM &  nRMSE & PSNR &  Memory & Parameters  \\
 &&&(dB)&(GB)&\\
\midrule

     End-to-End & 0.990(0.009) & 0.070(0.011) & 41.71(1.59)&10.5& 0.83M\\
     GLEAM & 0.993(0.006) & \textbf{0.057(0.014)} & \textbf{43.50(2.10)}&10.6& \textbf{2.50M}\\
     CS & 0.983(0.013) & 0.107(0.043) & 38.64(3.13)&N/A&N/A  \\
\bottomrule
\end{tabular}
}
\caption{\small \small{Reconstruction methods applied to the 3D unrolled PGD architecture at $R = 10$ on dynamic cardiac cine MRI. Comparison of image metrics, maximum GPU memory during training and number of learnable parameters for each method.}} 
\label{tbl:cine3D}
\end{table}

When P-GLEAM is used to distribute training on 2 GPUs, updates in individual modules can be computed in parallel, enabling faster training. Thus, P-GLEAM on 2 GPUs is faster by $1.6 \times$ compared to end-to-end learning, and GLEAM on 1 GPU. In this setting, it is not possible to compute $T_f$ and $T_b$ independently, since some forward and backward operations are done in parallel, without having to wait for each other.

\begin{figure*}[t!]
  \centering
  \begin{center}
      \includegraphics[width = 0.9\linewidth]{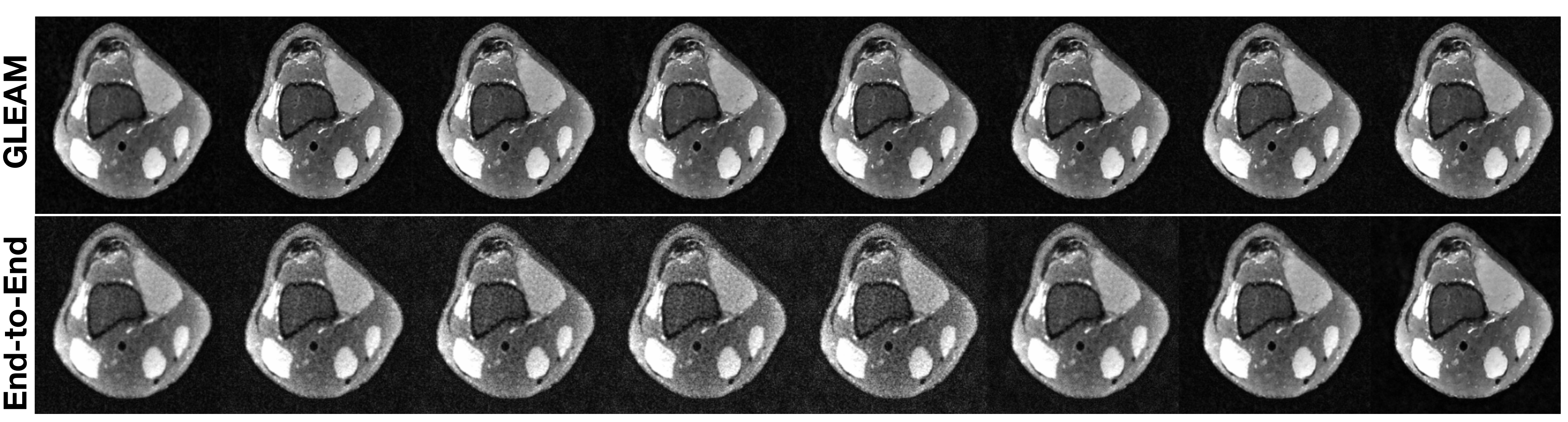}
  \end{center}
  \caption{\small{Interpretation of learned intermediate features during inference for each unrolled iteration $N_{inf}$. From left to right, images show inference run at $N_{inf} \in [8]$. End-to-end learning produces noisy intermediate features before a clean prediction is made at the end of the final layer. GLEAM iteratively refines its prediction at each unrolled iteration, mimicking a traditional iterative optimization algorithm. 
  }}
  \label{fig:features}
\end{figure*}

\begin{table}[t!]
\centering
 \resizebox{\columnwidth}{!}{
\begin{tabular}{llccccc}
\toprule

Method &      SSIM &          nRMSE &      PSNR  &Memory & Timing\\ 
& & & (dB) & (GB) & (iter/sec) \\
\midrule
     End-to-End & 0.993(0.005) & 0.059(0.015) & 43.33(2.15)&11.2& \textbf{4.26}\\
     Checkpointing & 0.993(0.005) & 0.059(0.016) & 43.35(2.19)&3.3& 5.75\\
     GLEAM & 0.993(0.006) & 0.061(0.017) & 42.97(2.25)&\textbf{3.1}& \textbf{4.29}\\
     CS & 0.983(0.013) & 0.107(0.043) & 38.64(3.13)&N/A&N/A \\
\bottomrule
\end{tabular}
}
\caption{\small{Reconstruction methods applied to the 3D MoDL architecture at $R = 10$ on dynamic cardiac cine MRI. Comparison of image metrics, maximum GPU memory during training and training time per iteration.}} 
\label{tbl:3d-modl}
\end{table} 

\vspace{-0.2cm}

\subsection{Interpreting Learned Reconstructions}
Although the final end-to-end networks for GLEAM and end-to-end BP have the same architecture, the progression of the learned intermediate features are different due to the different loss calculation paradigms that are employed. We can visualize these intermediate features by performing inference at different stages of the network. Here, we consider the networks trained via end-to-end BP and GLEAM from Table \ref{tbl:2d-modl} at $R = 16$, and run inference on each stage of the networks. Specifically, we run inference at $N_{inf} \in [8]$, where $N_{inf}$ denotes the number of unrolled steps are used at inference, and $[8]$ denotes the set of integers from $1$ to $8$. 

\begin{figure}[t!]
  \centering
  \begin{center}
      \includegraphics[width = 0.9\linewidth]{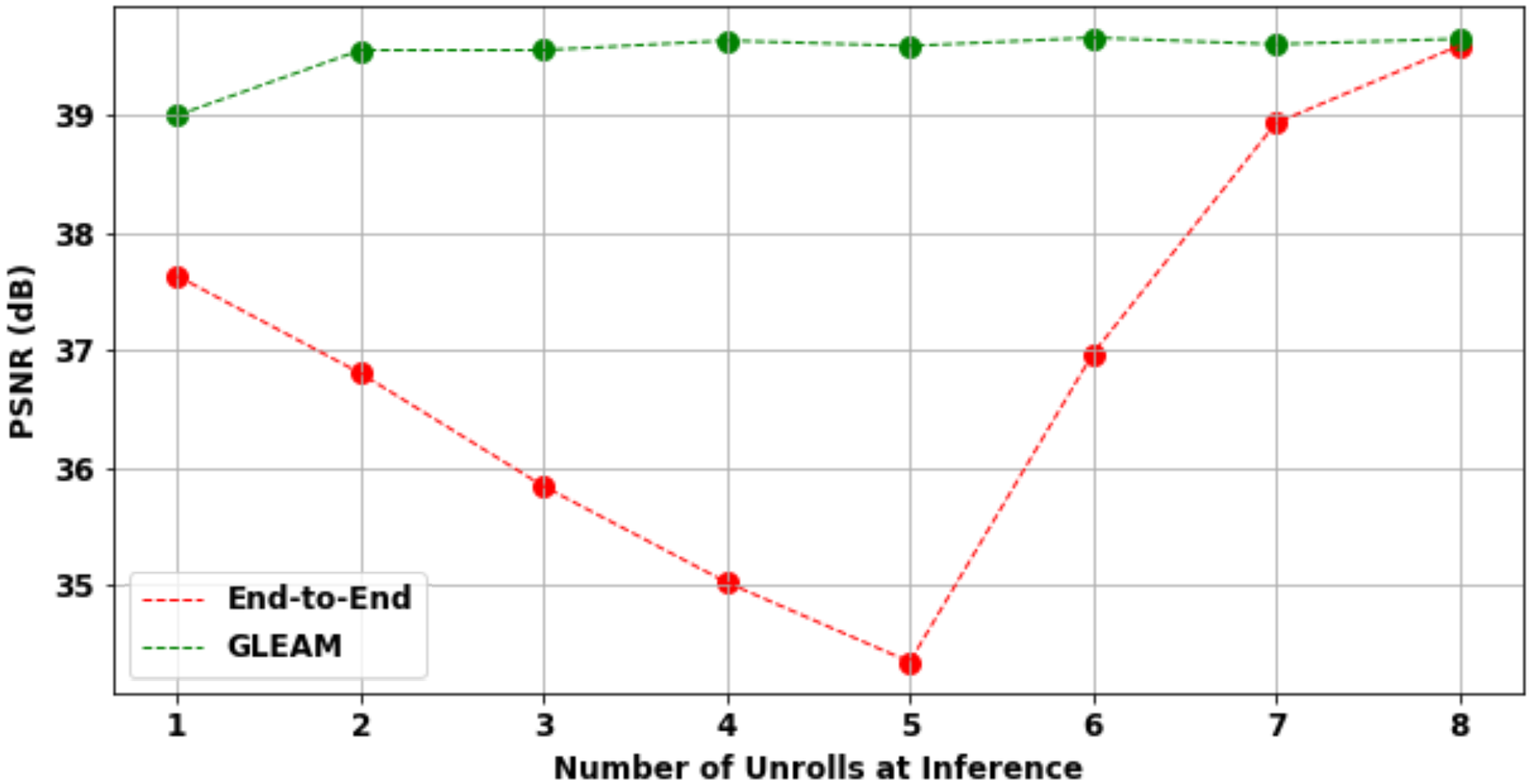}
  \end{center}
  \caption{\small{PSNR calculated on all test slices at each $N_{inf} \in [8]$. In end-to-end learning, PSNR is not monotonic, and attains its lowest value at $N_{inf} = 5$ whereas PSNR improves monotonically for GLEAM.}}
  \label{fig:inf-metrics}
\end{figure}

Fig. \ref{fig:features} shows intermediate features at the end of different stages $N_{inf}$ for each method, where Fig. \ref{fig:inf-metrics} shows PSNR calculated on all test slices at each $N_{inf} \in [8]$. In end-to-end BP, all modules are updated with an end-to-end loss, and the intermediate features are not controlled. As a result, they do not demonstrate iterative refinement of the features, and PSNR is not monotonic, attaining its lowest value at $N_{inf} = 5$. On the other hand, since each intermediate feature is learned to fit the ground-truth via a local loss in GLEAM, the features are refined at each iteration, and PSNR improves monotonically. The iterative refinement observed in the intermediate features of GLEAM resembles traditional iterative optimization algorithms used in CS \cite{cs-mri}.

\section{Discussion}
We presented GLEAM, a memory and time-efficient method for accelerated MRI reconstruction. In this section, we discuss key advantages, limitations, and future work.
\subsubsection{Advantages}
GLEAM reduces the memory requirements of unrolled networks during training without an increase in computation time. The improved memory and time requirements provided by GLEAM can enable the use of unrolled networks in challenging high-dimensional imaging scenarios. Furthermore, while we demonstrated our method on accelerated MRI reconstruction, it can be extended to an unrolled network used to solve any inverse problem that is constrained by memory. GLEAM also opens the opportunity to solve accelerated MRI reconstruction using only shallow networks, which are well studied and can provide more interpretable and robust solutions \cite{sahiner2020convex}.

In end-to-end learning of unrolled networks, the optimization algorithm has to be unrolled for a fixed number of iterations, and trained end-to-end. To determine the number of unrolls $N$ that is required for a specific application, the end-to-end network has to be trained from scratch for each $N$. In contrast, as we showed in Fig. \ref{fig:features}, once a network is trained via GLEAM for $N$ unrolled iterations, each sub-network in GLEAM with number of unrolled iterations $N_{inf} \in [N]$ has been optimized to predict the ground truth at the same time. Therefore, $N_{inf}$ can be selected at inference time, which obviates the need for extensive architecture search to tune $N$. 

The ability to perform inference at arbitrarily deep stages can provide benefits for inference time as well. As seen in Fig. \ref{fig:features}, performance might begin to saturate as $N_{inf}$ is increased. Thus, using GLEAM, the user can specify the desired $N_{inf}$ which might be smaller than $N$, to reduce the size of the end-to-end network at inference, and in turn speed up inference time. Such early exit schemes that use a portion of a trained network at inference have been deemed to be successful for image classification \cite{teerapittayanon2016branchynet}. Similar to ours, in a recent study, authors observed that using deep equilibrium models could enable computational budget to be selected at test time \cite{gilton2021deep}. Such methods can provide more efficient inference strategies compared to end-to-end learning.
 
 GLEAM provides flexibility during training. Each unrolled layer optimization can be tuned individually by selecting different loss functions, learning rate or other hyperparameters at different stages. GLEAM also enables users to specify $M$, which determines the memory usage during training. $M$ can be selected to adjust training memory for different applications. 

\subsubsection{Limitations}
A stated limitation of greedy learning is the decoupling of gradient updates, which prevents information flow between learned features among network modules that are optimized independently, theoretically resulting in less generalizable features as $M$ increases \cite{belilovsky2020decoupled}. In our experiments for this application, however, we found that GLEAM with $N = M$ generalizes as well as end-to-end learning, in addition to great improvements to memory consumption. In applications that may require a high number of unrolled iterations (e.g. $N > 100$), choosing $N = M$ might limit network expressivity, and further experimentation might be needed to find $M$ for good generalization.

\subsubsection{Future Work}
GLEAM could be extended to high-dimensional imaging scenarios such as DCE, 4D-flow, or 3D non-cartesian imaging, and to other learning settings such as semi-supervised \cite{desai2021noise2recon,desai2021vortex} or self-supervised \cite{Yaman_self} MRI.

\vspace{-0.4cm}

\section{Conclusion}
In this work, we presented GLEAM, an alternative to end-to-end BP for training unrolled networks via greedy learning. GLEAM allows a large reduction in the memory-footprint of training, while preserving the training time and generalization of BP, and enables parallelizing the greedy training objective to distribute training on multiple GPUs to speed up training. 
Experiments on multiple MRI datasets demonstrate the effectiveness of GLEAM in reducing memory during training while preserving training time and improving generalization.

\vspace{-0.35cm}

{\small
\bibliographystyle{IEEEtran}
\bibliography{greedy_arxiv}
}

\end{document}